\documentclass[preprint,12pt]{elsarticle}

\usepackage{amssymb}
\usepackage{float}
\usepackage{amsmath} 
\usepackage{amssymb}
\usepackage{geometry}
\usepackage{outlines}
\usepackage{lineno}
\usepackage{tabularx}
\DeclareMathOperator\arcsinh{arcsinh}

\biboptions{sort&compress}

\begin{document}

\begin{frontmatter}




\title{Nucleation of frictional slip: A yielding or a fracture process?}


\author[IfB]{Miguel Castellano}

\affiliation[IfB]{organization={Institute for Building Materials, ETH Zurich},country={Switzerland}}

\author[IfB]{Flavio Lorez}
\author[IfB]{David S. Kammer \corref{cor1}}
\ead{dkammer@ethz.ch}

\cortext[cor1]{Corresponding author}

\begin{abstract}
The onset of frictional sliding between contacting bodies under shear load is nucleated by the quasi-static growth of localized slip patches. After reaching a certain critical size, known as the nucleation length, these patches become unstable and continue growing dynamically, eventually causing the sliding of the entire interface. Two different theories have been used to compute the nucleation length of such patches depending on the dominant process driving their growth. If it is only the yielding of contact asperities (large-scale yielding), a stress criterion is applied, based on linear stability analysis, whereas if fracture dominates (small-scale yielding), an energy criterion is applied (Griffith's criterion), based on fracture mechanics and classical nucleation theory. Both approaches contain important underlying assumptions that are well-suited to describe either one situation or the other. However, what happens in-between is not captured by any of them. In this work, we use numerical simulations to study what is the dominant underlying process driving nucleation for different conditions of heterogeneity in the frictional strength of the interface and what are the implications for nucleation dynamics and the onset of frictional sliding. We show that large frictional heterogeneities enable a transition from a yielding-driven nucleation phase to a fracture-driven one. This transition occurs only above a certain level of heterogeneity and can either be quasi-static (stable) or dynamic (unstable), depending on the correlation length of frictional strength along the interface and the difference in strength between the strongest and the weakest point (the amplitude). Unstable transitions generate localized dynamic slip events, whose magnitude increases with higher correlation length and decreases with larger amplitude. Our work sheds new light on the role of heterogeneity and fracture in the nucleation of frictional slip, bridging the gap between the two main governing theories for nucleation. 
\end{abstract}

\begin{highlights}

\item Depending on heterogeneity, nucleation can be dominated by yielding or fracture.

\item The transition from yielding to fracture can produce frictional instabilities.

\item Magnitude increases with correlation length and decreases with amplitude of strength.

\item High heterogeneity favors stability and delays the onset of dynamic friction.

\end{highlights}

\begin{keyword}
friction \sep fracture \sep nucleation \sep slip \sep yielding \sep residual friction  
\end{keyword}

\end{frontmatter}


\section{Introduction}
\label{sec:intro}
A system of two bodies in frictional contact, subject to shear loading, accumulates stress on the contact asperities that form the interface. Typically, when loaded over their strength limit, these asperities start weakening, releasing some of the accumulated stress inelastically through the nucleation of frictional slip~\cite{brochardYieldFractureFailure2016}, and eventually break (or fracture), reaching a residual frictional strength. Collectively, the stress released by weakening asperities is transferred to stronger neighboring ones starting a chain reaction~\cite{geusHowCollectiveAsperity2019}. As a result, frictional slip nucleates locally and extends to neighboring areas creating sets of slip-patches. When these patches attain their critical size, either separately or by coalescence~\cite{scharNucleationFrictionalSliding2021}, they become unstable and grow dynamically until arrest. The post-nucleation processes have been shown to be well described by Linear Elastic Fracture Mechanics (LEFM). For instance, LEFM has proven very successful in determining the conditions for propagation and arrest of rupture fronts mediating the onset of dynamic frictional motion~\cite{kammerLinearElasticFracture2015,bayartFractureMechanicsDetermine2016,bayartRuptureDynamicsHeterogeneous2018}, including the stress fields near the front-tip~\cite{svetlizkyClassicalShearCracks2014,svetlizkyDynamicFieldsTip2020,rosakisRecentMilestonesUnraveling2020,giannakopoulosDynamicsFlexoelectricMaterials2020} or the propagation velocity of such fronts~\cite{svetlizkyPropertiesShearStress2016,svetlizkyBrittleFractureTheory2017}. However, it remains unclear whether or not LEFM can describe the nucleation of frictional sliding. \par

The nucleation process, characterized by the stable quasi-static growth of slip-patches along the frictional interface, can typically undergo two different phases~\cite{dieterichPreseismicFaultSlip1978,ohnakaCharacteristicFeaturesLocal1990,ohnakaScalingShearRupture1999,rubinEarthquakeNucleationAging2005,mclaskeyForeshocksNucleationStickslip2013,kanekoOnsetLaboratoryEarthquakes2016,mclaskeyEarthquakeInitiationLaboratory2019}; a yielding phase, characterized by the weakening of frictional strength with slip, and a fracture phase, where the strength remains relatively constant with slip (residual friction). Depending on the governing phase driving the growth of the nucleating patch, LEFM will or will not be applicable. This is because one of the central assumptions of LEFM is small-scale yielding, which means that the length-scale associated to yielding should be small compared to the overall size of the slip patch~\cite{uenishiUniversalNucleationLength2003,dieterichEarthquakeNucleationFaults1992,campilloInitiationAntiplaneShear1997}. \par 

Previous numerical works on nucleation of frictional sliding have mostly focused on the yielding phase, either because residual friction is disregarded~\cite{uenishiUniversalNucleationLength2003} or simply because it is not triggered during nucleation~\cite{albertiniStochasticPropertiesStatic2021,lebihainEarthquakeNucleationFaults2021}. Other works that do trigger residual friction concentrate mainly on the effect of background stress and pore pressure~\cite{viescaNucleationSlipweakeningRupture2012a,garagashNucleationArrestDynamic2012,brantutEarthquakeNucleationIntact2015,azadNucleationDynamicSlip2017}, while some others explore the role of heterogeneity but are still limited to a single source~\cite{ampueroPropertiesDynamicEarthquake2006, rippergerEarthquakeSourceCharacteristics2007,hillersStatisticalPropertiesSeismicity2007} or focus exclusively on the quasi-static features of nucleation~\cite{rubinEarthquakeNucleationAging2005,ampueroEarthquakeNucleationRate2008}, missing out on the rich dynamics of late nucleation mechanisms triggered by the onset of fracture. However, there is experimental~\cite{ohnakaEarthquakeSourceNucleation1992,mclaskeyForeshocksNucleationStickslip2013,latourCharacterizationNucleationLaboratory2013,harbordEarthquakeNucleationRough2017} and numerical evidence~\cite{bar-sinaiInstabilitiesFrictionalInterfaces2013,dublanchetDynamicsEarthquakePrecursors2018,cattaniaPrecursorySlowSlip2021} suggesting that these mechanisms are actually decisive in determining the onset of frictional motion. For instance, recent experiments by Gvirtzman and Fineberg~\cite{gvirtzmanNucleationFrontsIgnite2021} show that slow nucleation fronts prepare the stage for the onset of rapid dynamic rupture in laboratory frictional interfaces, while similar ones by McLaskey~\cite{mclaskeyEarthquakeInitiationLaboratory2019} show the importance of foreshock activity during nucleation. \par 


Here, we study the effect of heterogeneity on the stability of both nucleation phases for both single-patch and stochastic nucleation set-ups. Using elastostatic solutions and dynamic simulations, we show that the correlation length and the amplitude of the frictional strength along the interface control the onset of the fracture phase stability, producing three different nucleation regimes characterized by the relative dominance between the yielding and the fracture phase. These regimes are observed for both slip-weakening and rate-and-state friction laws. Our observations are in good agreement with a range of phenomena described in the literature, both in simulations and experiments~\cite{rubinsteinDetachmentFrontsOnset2004,ben-davidSlipstickEvolutionFrictional2010, tromborgSlowSlipTransition2014,gvirtzmanNucleationFrontsIgnite2021,cebryRoleBackgroundStress2022b}, and provide a unifying framework for nucleation of frictional sliding.

The paper is organized as follows. In section~\ref{sec:theory}, we introduce theoretical concepts regarding the nucleation of frictional slip. In section~\ref{sec:single-patch}, we study the nucleation mechanisms of a single nucleation patch and in section~\ref{sec:stochastic}, we extend our analysis to stochastic frictional interfaces. 


\section{Theory for the nucleation of frictional slip}
\label{sec:theory}

\subsection{Elastic equations}
\label{sec:theory_elastic_equations}
Frictional slip is controlled by the shear stress dynamics along the interface, which can be described by a set of governing elastodynamic equations. In particular, the fully-dynamic formulation of these equations accounts for all dynamic effects, including the wave-mediated stress transfers. These are computed through a convolution integral over the causality cone of distant points~\cite{geubelleSpectralMethodThreedimensional1995} and introduce long-range interactions into the system, which may have an important effect on stability, as observed for the dynamics of elastic depinning ~\cite{tanguyIndividualCollectivePinning1998,degeusScalingTheoryStatistics2022,geusHowCollectiveAsperity2019}. This formulation is given as follows:

\begin{equation}
   \tau_0(x,t) + f(x,t)  = \tau_{\mathrm{f}}(x,t) + \frac{\mu^{\prime}}{2c_s}V(x,t),
   \label{eq:fully-dynamic}
\end{equation} 
 where $\mu^{\prime} =  \mu/(1-\nu)$ for modes I and II, $\mu$ being the shear modulus, $\nu$ is the Poisson's ratio, $c_s$ the shear wave speed and $V$ the slip rate. Here, $\tau_\mathrm{f}(x,t)$ is the frictional strength, described by a constitutive law, $\tau_0(x,t)$ the external load, $f(x,t)$ is a convolution integral containing both the quasi-static and wave-mediated stress transfers and the last term is the radiation damping term, which accounts for the local dynamic release of stress. \par
 
Quasi-static models ignore inertial effects and describe only the static equilibrium, given by
\begin{equation}
     \tau_{\mathrm{f}}(x,t) =  \tau_0(x,t) - \frac{\mu^{\prime}}{2\pi}\int_{\ell_-(t)}^{\ell_+(t)} \frac{\partial \delta(\xi,t)/\partial \xi}{x - \xi} d\xi,
     \label{eq:quasi-static}
 \end{equation}
 where the last term, valid for plane-strain conditions~\cite{bilby19868fracture}, accounts for the quasi-static stress transfer due to the weakening of asperities over a given slip-patch going from $x = \ell_-(t)$ to $x = \ell_+(t)$, the length of the patch is given by $\ell(t) = \ell_+(t) - \ell_-(t) $ and slip is $\delta(\xi,t)$. \par

The effect of slip $\delta$ on the frictional strength $\tau_{\mathrm{f}}(x,t)$ is captured by the constitutive law, which can take many forms and plays a major role in controlling the stability of the interface. 
  
\subsection{Constitutive laws}
\label{sec:theory_const_laws}
In our work,  we use two different constitutive models, which are among the most frequently used in the literature: A simple linear slip-weakening friction law~\cite{idaCohesiveForceTip1972}, which is instrumental for a better understanding of the underlying mechanics of the system, and a regularized version of the more realistic rate-and-state law~\cite{riceSlipComplexityEarthquake1996}.\par 

According to the linear slip-weakening law, frictional strength $\tau_{\mathrm{f}}(x)$ is modelled as
\begin{equation}
\tau_{\mathrm{f}}(\delta,x) =  
    \begin{cases}
      \tau_p(x) -W\delta(x), & \text{if}\ \delta(x) < \delta_c (x) \\
      \tau_{r}, & \text{if}\ \delta(x) \geq \delta_c (x)
    \end{cases},
    \label{eq:slip-weakening}
\end{equation}
where $W = (\tau_p(x) - \tau_{r})/\delta_c (x)$ is the weakening rate, $\tau_p(x)$ is the local peak strength, $\tau_{r}$ the residual strength, $\delta(x)$ the slip computed as the local relative tangential displacement at the interface, and $\delta_c (x)$ is its critical value. Here, the toughness is coupled to the strength, which is a realistic assumption for friction, assuming dependence on normal stress~\cite{bayartFractureMechanicsDetermine2016}. \par

The rate-and-state dependent model, in contrast, is a regularized version~\cite{riceSlipComplexityEarthquake1996} of the original one proposed by Dieterich~\cite{dieterichModelingRockFriction1979} and Ruina~\cite{ruinaSlipInstabilityState1983}, which writes
\begin{equation}
    \tau_{\mathrm{f}}(V,\theta) = a \sigma 
    \arcsinh \left[ \frac{V}{2V^*} \exp \left( \frac{\Psi}{a}\right) \right],
    \label{eq:rate-and-state}
\end{equation}
where $\Psi = f^* + b\ln(V^* \theta / D_c)$, $V =\dot{\delta}$, $\sigma$ is the contact pressure, $\mathrm{f}^*$ and $V^*$ are reference values for the friction coefficient and slip velocity, $D_c$ a characteristic slip distance (distinct from $\delta_c$), $a$ and $b$ two constitutive parameters and $\theta$ is the state variable, which represents the local contact area of asperities, expressed in time-units as a proxy for the asperity 'life-span'. It follows an aging law evolution, given by
\begin{equation}
    \dot{\theta} = 1 - \frac{V\theta}{D_c}.
    \label{eq:evolution-law}
\end{equation}

With the elastic equation and the constitutive law for the frictional strength, we can compute the nucleation length, which is the maximum size that a slipping patch can reach before becoming unstable. 
 
\subsection{Theoretical nucleation length}
\label{sec:theory_nuc_length}
We consider two independent approaches for the nucleation length $\ell_c$. First, we summarize a stress-based criterion, which is relevant for situations where the small-scale yielding assumption of LEFM is not satisfied. Then, we present an energetic approach, based on LEFM, which is applicable to interfaces containing slip patches characterized mainly by residual friction~\cite{cornettiFiniteFractureMechanics2006}.

\subsubsection{Stress criterion}
\label{sec:theory_nuc_length_uenishi}
Linear stability analysis (LSA) provides a powerful tool to compute the maximum length of a stable slip-patch in a given frictional interface. Using LSA, a stability criterion can be established for rate-and-state friction~\cite{ruinaSlipInstabilityState1983}, which sets the critical stiffness of a spring-block slider to be $K_c = \sigma (b-a)/D_c$. According to this criterion, the nucleation length is
\begin{equation}
    \ell_c \equiv \ell_{\tau}^{RS} = C \frac{\mu^{\prime}D_c}{\sigma (b-a)},
    \label{eq:RS-length}
\end{equation} 
also referred to as $L_{b-a}$ in the literature, where $C$ is a dimensionless shape factor~\cite{rubinEarthquakeNucleationAging2005,harbordEarthquakeNucleationRough2017}.

Similarly, for slip-weakening friction, we can use LSA to compute the universal nucleation length for any non-uniform loading condition~\cite{campilloInitiationAntiplaneShear1997,uenishiUniversalNucleationLength2003,scharNucleationFrictionalSliding2021,albertiniStochasticPropertiesStatic2021,lebihainEarthquakeNucleationFaults2021}. Uenishi and Rice~\cite{uenishiUniversalNucleationLength2003} showed that starting from the equation for quasi-static elastic equilibrium (Eq.~\ref{eq:quasi-static}) and assuming a constant weakening rate $W$, one can derive an eigenvalue equation that is used to find the smallest positive eigenvalue, which satisfies the condition for instability. This leads to the nucleation length
\begin{equation}
    \ell_c \equiv \ell_{\tau}^{SW} = 1.158 \; \frac{\mu^{\prime}}{W},
    \label{eq:SW-length}
\end{equation}
which depends only on the weakening rate $W$ and assumes that no asperity within the yielding patch has reached its critical slip. Hence, it cannot be applied to interfaces containing patches of residual strength. In such cases, an energy criterion is applied. 

\subsubsection{Energy criterion (Griffith's criterion for nucleation)}
\label{sec:theory_nuc_length_griffith}
Classical nucleation theory (CNT) states that a phase will grow whenever this growth introduces a reduction of its free energy. Under the assumption of small-scale yielding, fracture mechanics uses CNT to introduce a stability criterion based on an energy balance to determine the nucleation size of a crack. This criterion establishes that a crack will become unstable whenever the energy release rate $G$, which is the strain energy released per unit of crack growth, will match locally the fracture energy $\Gamma$ of the interface ($G = \Gamma$). \par 

For an existing shear crack of length $\ell$, the static energy release rate under small-scale yielding conditions is given by 
\begin{equation}
    G_s(\ell) = \frac{K_{II}(\ell)^2}{E^{\prime}},
    \label{eq:ERR}
\end{equation}
where $E^{\prime} = E/(1-\nu^2)$. $K_{II}(\ell) = \Delta \tau_{0r} \sqrt{\pi (\ell - \mathcal{X}_c) / 2}$ represents the stress intensity factor of a central mode-II (shear) crack subject to a remote and uniform shear load $\tau_0$, where $\Delta \tau_{0r} = \tau_0 - \tau_r$ is the stress-drop and $\mathcal{X}_c$ is the size of the cohesive zone. This approach directly considers a cohesive crack, \textit{i.e.}, a crack with a cohesive zone, but it converges to the singular crack for $\mathcal{X}_c \rightarrow 0$. \par 

The fracture energy, instead, is often a fixed property of the interface accounting for the energy that needs to be invested for an existing crack to grow, and for linear-slip weakening friction, it is given by
\begin{equation}
    \Gamma = \frac{\Delta \tau_{pr}^2}{2W},
    \label{eq:fracture-energy}
\end{equation}
where $\Delta \tau_{pr} = \tau_p - \tau_r$ is the peak-to-residual stress-drop. \par 
However, in the case of an intact interface, without pre-existing cracks, the fracture energy is replaced by the more general crack-growth resistance $R_c$ to account for the build-up of the cohesive zone, which depends on the crack length $\ell$ as  
\begin{equation}
    R_c(\ell,t) = \int_{\ell- \mathcal{X}_c}^{\ell } \tau(x,t) \frac{\partial \delta(x,t)}{\partial x} dx,
    \label{eq:R_c}
\end{equation}
which is applicable to any constitute friction law.
Most materials have a rising R-curve, meaning that $R_c$ grows with crack length until reaching its steady state value, which is equal to the fracture energy $\Gamma$. In such cases, the process zone or cohesive zone that forms at the crack-tip provides further stability to the crack and adds an extra condition for instability, which depends on the rate at which $G_s$ and $R_c$ evolve. Therefore, the nucleation length $\ell_c$ for a cohesive crack growing in an intact interface is given by two conditions, which are
\noindent 
\begin{subequations}

\label{eq:energy-balance}
\begin{minipage}[t]{0.3\textwidth}
\vspace{-0.3cm}
\begin{equation}
        \label{eq:GcRc}
        G_s (\ell_c) = R_c (\ell_c)
\end{equation}

\end{minipage}\hfill
\begin{minipage}[t]{0.5\textwidth}\begin{equation}
   \label{eq:dGCdRc}
        \frac{\partial G_s(\ell)}{ \partial \ell} \bigg \rvert_{\ell = \ell_c} = \frac{\partial R_c(\ell)}{ \partial \ell} \bigg \rvert_{\ell=\ell_c}
\end{equation}\end{minipage}

\end{subequations}

It is worth mentioning that closed-form solutions for $\ell_c$ are often missing when $\Gamma(x)$ and/or $\Delta \tau_{0r}(x,t)$ are non-uniform. This is particularly true when $\Gamma = R_c$, which is often not known a priori. For this reason, the textbook Griffith's length, which assumes uniform $\Gamma$ and $\Delta \tau_{0r}$, is often used as reference. For slip-weakening friction, this solution is given by
\begin{equation}
    \ell_c \equiv \ell_{G}^{SW} = \frac{4\mu^{\prime}\Gamma}{\pi \Delta \tau_{0r}^2}.
    \label{eq:SW-crack-length}
\end{equation}
and the equivalent solution for rate-and-state friction with aging, which is also referred to as $L_{\infty}$, is given by
\begin{equation}
    \ell_c \equiv \ell_{G}^{RS} = \frac{1}{\pi} \left( \frac{b}{b-a}\right)^2 L_b
\label{eq:RS-crack-length}
\end{equation}
where $b \sim \Delta \tau_{pr}$, $(b-a) \sim \Delta \tau_{0r}$, $a/b > 0.5$ and $L_b = \mu^{\prime}D_c / b \sigma$. The ratio $a/b$ determines whether Eq.~\ref{eq:RS-length} or Eq.~\ref{eq:RS-crack-length} is relevant for nucleation in a given system~\cite{rubinEarthquakeNucleationAging2005}. \par

In summary, we have reviewed the elastodynamic equations governing the interface, the constitutive laws that model the frictional response of asperities and the different criteria for the nucleation length of slip patches, which depend on the governing phase driving their growth (yielding or fracture). However, it is still unclear what happens during nucleation when both phases become relevant. For this reason, we first study the nucleation of a single patch over different interface configurations and eventually consider a more realistic stochastic scenario. \par 

\section{Single patch nucleation}
\label{sec:single-patch}

In a first step, we consider the nucleation of a single slip patch from an isolated weak zone embedded in a stronger domain. We describe this non-uniformity using an inverted Gaussian function
\begin{equation}
    q(x) = q^{max}\left[ 1 - A\exp(\frac{-x^2}{2\eta^2})\right],
    \label{eq:qprofile}
\end{equation}
where $q^{max}$ is the maximum value of $q(x)$, $A$ represents the amplitude of the heterogeneity and $\eta = \lambda/2\sqrt{2\ln{2}}$, where $\lambda$ denotes the Full Width at Half Maximum (FWHM) of the Gaussian, a surrogate for the correlation length of the profile. This setup allows us to study the nucleation process in isolation, avoiding coalescence and other potential interactions between neighboring patches~\cite{scharNucleationFrictionalSliding2021}. \par  

\begin{figure}
    \centering
    \centerline{\includegraphics[scale=1.25]{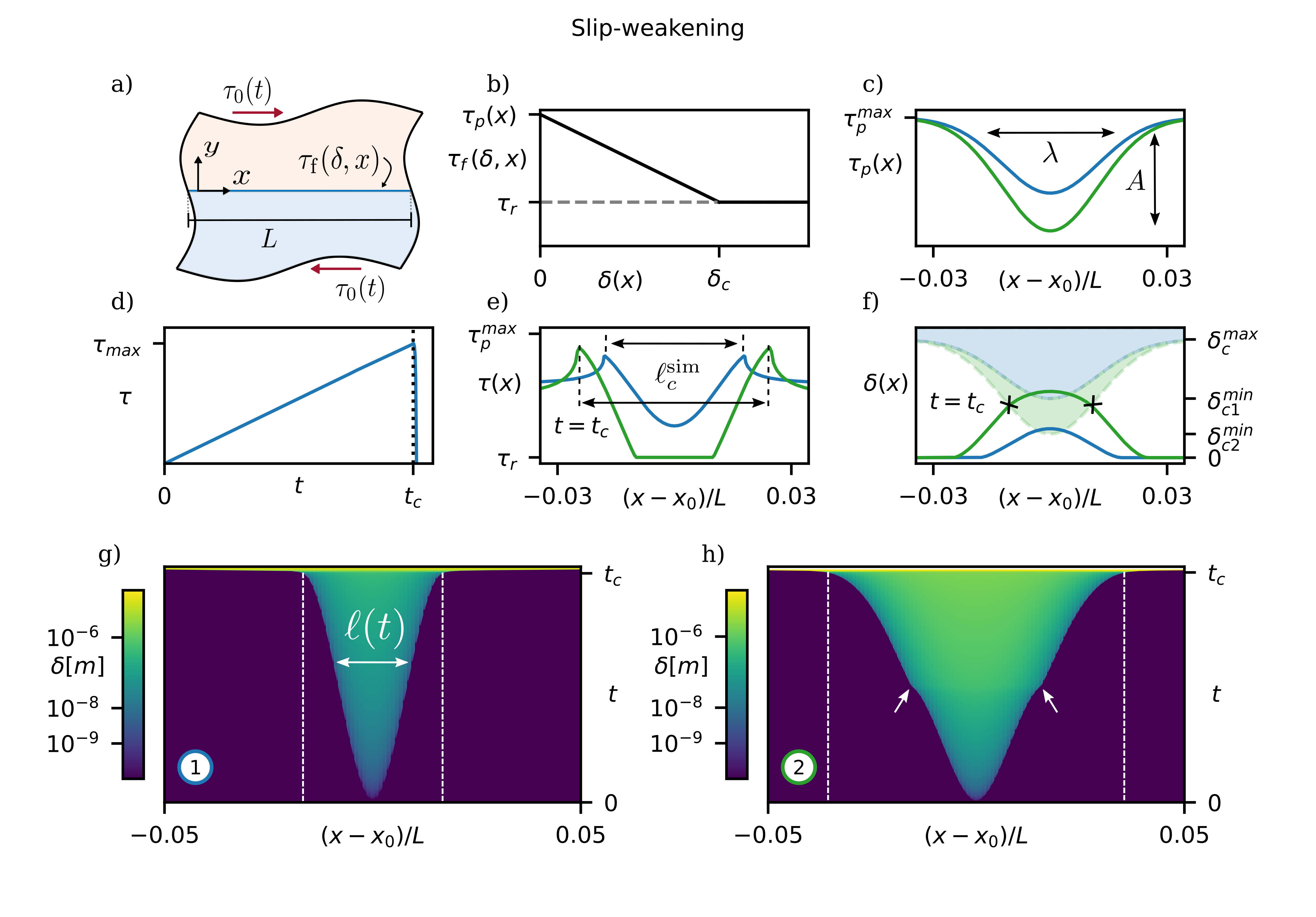}}
    \caption{Simulations of the onset of frictional sliding for a single nucleation patch governed by linear slip-weakening friction. a) Illustration of the simulation set-up. $\tau_0$ is the applied shear load, $L$ the total length of the interface and $\tau_\mathrm{f}(\delta,x)$ the frictional strength along the interface. b) Linear slip-weakening constitutive law.  c) Distribution of peak-strength.  d) Integrated macroscopic frictional stress over time. The critical nucleation time $t_c$ is determined as the point of limiting friction, where the stress drops abruptly. e) Shear stress distribution at the critical nucleation time. The nucleation lengths measured from the blue and green simulations $\ell_{c}^{sim}$ are indicated by black double-arrows. f) Slip accumulated along the interface at critical nucleation time. The colored area marks the onset of residual friction ($\delta > \delta_c(x)$) for each system.  g,h) Slip evolution in space and time. $\ell$ is the length of the nucleation patch. Dashed white lines indicate $\ell_{c}^{sim}$. All sub-figures show data from the same two simulations. Number 1 (shown in blue and in g) showcases the large-scale yielding regime, with $A=0.5$ and $\lambda/\ell_{\tau}^{SW}= 0.7$ and number 2 (green and in f), the small-scale yielding regime (static), with $A=0.75$ and $\lambda/\ell_{\tau}^{SW} = 0.7$.}
    \label{fig:SW}
\end{figure}

\subsection{Slip-weakening friction}

We consider an interface between two semi-infinite elastic half-spaces of length $L$, as depicted in Fig.~\ref{fig:SW}a (parameter values for simulation are provided in \ref{sec:simparams}). The interface is governed by linear slip-weakening friction (see Fig.~\ref{fig:SW}b and Eq.~\ref{eq:slip-weakening}) with a non-uniform strength profile $\tau_p(x) = q(x)$. Two representative examples are shown in Fig.~\ref{fig:SW}c with different amplitudes $A$. The interface is loaded with a uniform load $\tau_0(t)$ that increases quasi-statically over time. Simulations are run using the Spectral Boundary Integral (SBI) Method~\cite{geubelleSpectralMethodThreedimensional1995, breitenfeldNumericalAnalysisDynamic1998,kammerUGUCASpectralboundaryintegralMethod2021}, which solves the fully dynamic formulation of the elastodynamic equations in the spectral domain. \par 

In our simulations, we observe the following behavior. When the load level reaches the peak-strength of asperities in the central weaker zone, these start yielding, nucleating a slowly growing area of local slip of size $\ell(t)$ (see Fig.~\ref{fig:SW}g). As they weaken, excess stress is transferred to surrounding asperities on the edge of the growing patch, pushing it to grow, while slip accumulates inside. At a critical time $t = t_c$, the macroscopic frictional strength of the interface drops drastically, indicating that the interface has failed (Fig.~\ref{fig:SW}d). This is due to the slip patch reaching the nucleation length $\ell_{c}^{\textrm{sim}}$, becoming unstable and propagating rapidly along the interface (see yellow area at the very top in Fig.~\ref{fig:SW}g). In both shown examples (blue and green in Fig.~\ref{fig:SW}c) this process looks relatively similar. However, there are crucial differences. \par

Considering first the configuration with lower amplitude (blue in Fig.~\ref{fig:SW}c), we note that at $t = t_c$, no asperity has reached its critical slip (Fig.~\ref{fig:SW}f), and therefore the residual strength has not been reached (Fig.~\ref{fig:SW}e). This means that the nucleation has not yet entered the fracture phase and can be described by the stress criterion. Below we will show that $\ell_{\tau}^{SW}$ describes quantitatively well the nucleation length $\ell_c^{\textrm{sim}}$ obtained from the simulations. Considering now the case of the interface with higher amplitude (green in Fig.~\ref{fig:SW}c), we note that the asperities in the weak zone, which are much weaker than their neighbors, reach their critical slip $\delta_c(x)$ (Fig.~\ref{fig:SW}f) and break, \textit{i.e.}, $\tau_{\mathrm{f}}(x) = \tau_r$), (Figs~\ref{fig:SW}e) at $t<t_c$. Yet, this does not trigger any instability, since the stronger neighboring asperities continue to bear the stress released by broken asperities, containing the propagation of fracture. Instead, a stable cohesive shear-crack is formed, which continues to grow quasi-statically as the interface is loaded until it reaches its nucleation length described by the energy criterion. This transition from a yielding patch to a stable cohesive crack is quasi-static but can still be perceived on Fig.~\ref{fig:SW}h as a little kink in the growth of the slip patch (see white arrows). \par 

\begin{figure}
    \centering
    \includegraphics[scale=1.25]{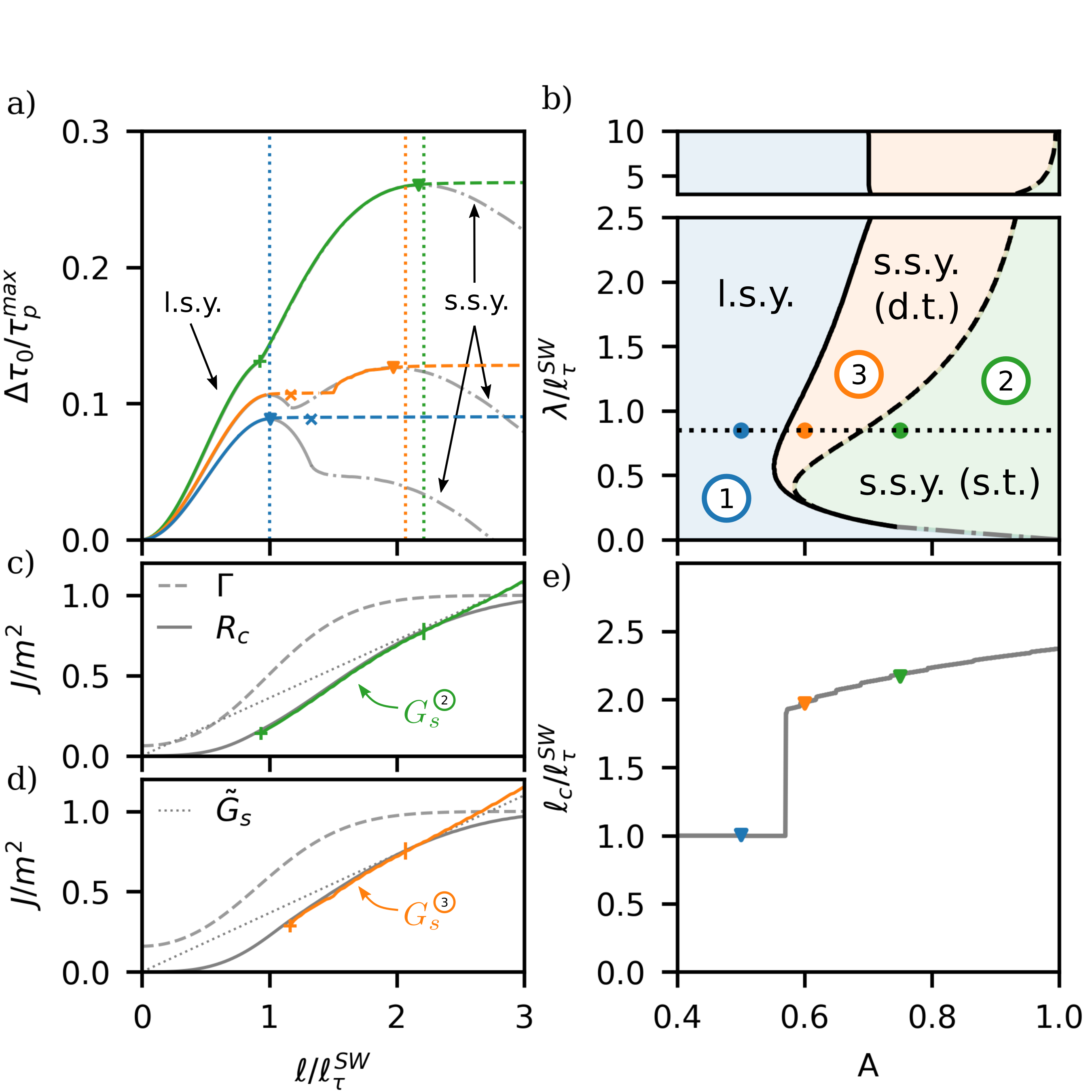}
    \caption{Nucleation regimes for a single nucleation patch governed by slip-weakening friction. \textbf{a)} Normalized load increment applied on the interface $\Delta \tau_0 = \tau_0(t) - \tau_p^{min}$, where $\tau_p^{min} = \tau_p(x_0)$, as a function of the normalized slip-patch length. Gray curves are the elastostatic solutions $\Delta \tau_0^s(\ell)$, where the small-scale yielding parts are dash-dotted. Colored lines are drawn from dynamic simulations where dashed segments represent unstable patch growth. The '$\times$' signs mark the onset of residual friction, and the triangles, the ultimate failure point, defining $\ell_c^{\textrm{sim}}$. The vertical blue dotted line represents the nucleation length as computed with Eq.~\ref{eq:SW-length}. The vertical orange and green dotted lines are computed from an approximate energy-criterion based on Eq.~\ref{eq:energy-balance} and \ref{eq:ERRapprox}. \textbf{b)} Phase diagram showing the resulting nucleation regimes for every interface configuration ($\lambda$,$A$). The colored dots indicate the parameters of the simulations shown in 'a)', with $A=0.5$ (blue), $A=0.6$ (orange) and $A=0.75$ (green), and $\lambda/\ell_{\tau}^{SW} = 0.85$. The gray dashed-dotted line is a linear extrapolation of the boundary.   \textbf{c,d)} R-curve $R_c(\ell)$ (solid gray), $\Gamma(x)$ (dashed gray) and $G_s(\ell)$ (in color) for the green and orange cases, respectively, displayed in 'a)'. The dotted gray tangent lines are the approximate $\tilde{G}_s(\ell)$ given by Eq.~\ref{eq:ERRapprox} that satisfy Eqs.~\ref{eq:energy-balance} and hence set the nucleation length, marked in 'a)' as vertical dotted lines. \textbf{e)} Normalized critical length, computed from the elastostatic solutions, as a function of $A$ for $\lambda / \ell_{\tau}^{SW} = 0.85$. Colored triangles refer to the simulations in 'a)'.}
    \label{fig:SW-regimes}
\end{figure}

The nucleation process is best illustrated on Fig.~\ref{fig:SW-regimes}a (see gray curves), where we compute the elastostatic solutions (Eq.~\ref{eq:quasi-static}) of the load increment as a function of the slip patch length $\Delta \tau_0^s(\ell)$ for three different cases using the piece-wise constant slip method from Garagash and Germanovich, 2012~\cite{garagashNucleationArrestDynamic2012} (see also \ref{sec:slipmethod}). Additionally, we plot the dynamic solutions $\Delta \tau_0^d(\ell)$ obtained from our SBI simulations (see colored curves). These solutions can be divided into a large-scale yielding part (before the '$\times$' on Fig.~\ref{fig:SW-regimes}a, which marks the onset of fracture) and a small-scale yielding part (after the '$\times$', where fracture takes over). First, we consider simulation 1 from Fig.~\ref{fig:SW} (shown in blue). In Fig.~\ref{fig:SW-regimes}a, we observe that the patch grows quasi-statically (following the static solution perfectly) until reaching $\ell_{\tau}^{SW}$, where $\Delta \tau_0^s(\ell)$ reaches its maximum. From this point on, the slip patch becomes unstable and propagates dynamically (indicated by a dashed line). Since the dynamic simulations are load-controlled through quasi-static loading, the load can only increase, and therefore, only the increasing parts of $\Delta \tau_0^s(\ell)$ are actually stable, which explains the onset of instability. Hence, simulation 1 is, quantitatively well described by the stress-based criterion and therefore, $\ell_c^{\textrm{sim}} =\ell_{\tau}^{SW}$ (in Fig.~\ref{fig:SW-regimes}a compare blue triangle indicating $\ell_c^{\textrm{sim}}$ with dashed line corresponding to $\ell_{\tau}^{SW}$). As the final point of instability is given by the maximum of the large-scale yielding part of $\Delta \tau_0^s(\ell)$, we call this the large-scale yielding regime (l.s.y.).  \par

Simulation 2 from Fig.~\ref{fig:SW} (shown in green), however, presents stable quasi-static crack growth far beyond $\ell_{\tau}^{SW}$ (see Fig.~\ref{fig:SW-regimes}a). This is because the slip patch reaches the fracture phase at $\ell<\ell_{\tau}^{SW}$ (marked by '$\times$' on Fig.~\ref{fig:SW-regimes}a) and continues growing as a stable cohesive crack. Past this point, as discussed in Sec.~\ref{sec:intro}, crack growth is governed by an energy balance given by Eq.~\ref{eq:GcRc} (see Fig.~\ref{fig:SW-regimes}c), which assumes small-scale yielding conditions. Eventually, the crack becomes unstable when $R_c(\ell) > G_s (\ell)$ (marked by vertical tick in Fig.~\ref{fig:SW-regimes}c), which appears to be in good quantitative comparison with the small-scale yielding part of the elastostatic solution (see triangle in Fig.~\ref{fig:SW-regimes}a). Accordingly, this is the small-scale yielding regime (s.s.y.) with static transition (s.t.), to account for the quasi-static transition from the yielding to the fracture phase (below we will discuss also a different type of transition).

These results have demonstrated that slip patch growth in the s.s.y. regime is well described by the energy balance. However, a specific nucleation criterion remains missing. As discussed in Sec.~\ref{sec:theory_nuc_length}, Griffith's solution (Eq.~\ref{eq:SW-crack-length}) is not directly applicable because the critical stress level is unknown. However, if we relax the assumption of uniform $\Gamma$ and use the actual $R_c(\ell)$, and approximate the energy release rate given by Eq.~\ref{eq:ERR} as
\begin{equation}
    \tilde{G}_s(\ell,\Delta \tau_{0r}) \approx \frac{\pi \Delta \tau_{0r}^2 \ell}{2E^{\prime}},
    \label{eq:ERRapprox}
\end{equation}
which assumes uniform $\Delta \tau_{0r}$ and $\mathcal{X}_c = 0$ (see dotted line in Fig.~\ref{fig:SW-regimes}c), we can solve Eqs.~\ref{eq:energy-balance} and find a quantitatively good prediction of the nucleation length (see vertical dotted line in Fig.~\ref{fig:SW-regimes}a). Finally, we should note that if we had entirely neglected the cohesive zone and used $\Gamma(\ell)$ instead of $R_c(\ell)$, the prediction for the critical nucleation length would be considerably worse (imagine a $G_s(\ell) \sim \ell$ tangent to $\Gamma(\ell)$ in Fig.~\ref{fig:SW-regimes}c).

So far, our results have demonstrated that the stress-based and energy-based nucleation criteria work well in describing the instability onset of quasi-static slip patches in distinct parts of the parameter space (width and depth) of a simple weak zone. Our simulations, however, unveiled a third nucleation regime. 
In this case (denoted simulation 3), similar to simulation 1, the slip patch first grows slowly until it reaches $\ell_{\tau}^{SW}$, when it becomes unstable and starts growing dynamically (see orange in Fig.~\ref{fig:SW-regimes}a). However, shortly after becoming unstable, it turns into a cohesive crack (marked by '$\times$' in Fig.~\ref{fig:SW-regimes}a), and finds a new static solution (the small-scale yielding one), which leads to arrest of the slip patch.  Beyond this point, additional loading will lead to further stable crack growth until the energy criterion is satisfied (Fig.~\ref{fig:SW-regimes}d). Although the eventual (and final) failure is governed by the same criterion as simulation 2 (green), for simulation 3 the transition from yielding to fracture is unstable and, therefore, dynamic. For that reason, we name this the small-scale yielding regime (s.s.y.) with dynamic transition (d.t.), to differentiate it from the previous. \par

Next, we explore the parametric space of the three nucleation regimes for frictional sliding at a simple weak patch (see Fig.~\ref{fig:SW-regimes}b). 
First, we observe that strength profiles with amplitudes below a critical value $A < \approx 0.5$ always lead to a large-scale yielding nucleation independent of the spatial characteristics of the profile. Above this limit, however, all three nucleation regimes exist and occupy different parts of the parametric space. For $\lambda / \ell_{\tau}^{SW} \rightarrow 0$, nucleation tends to be dominated solely by the large-scale yielding regime independent of $A$, and for $\lambda / \ell_{\tau}^{SW} \gg 1$ nucleation mainly occurs either through an l.s.y. or an s.s.y. (d.t.) regime for small and large $A$, respectively. Accordingly, the s.s.y (s.t.) regime occurs only for intermediate $\lambda / \ell_{\tau}^{SW} \sim 1$ at very high amplitude $A$. 
Finally, we note that the existence of different nucleation regimes leads to varying critical nucleation length (see Fig.~\ref{fig:SW-regimes}e). At low $A$, the critical length is constant for varying $A$, and is given by $\ell_c = \ell_{\tau}^{SW}$. At the point of change in nucleation regime, $\ell_c$ presents a discontinuity and beyond that grows as a function of $A$ (at constant $\zeta$), \textit{i.e.}, $\ell_c = \ell_c(A)$.

\begin{figure}
    \centering
    \centerline{\includegraphics[scale=1.25]{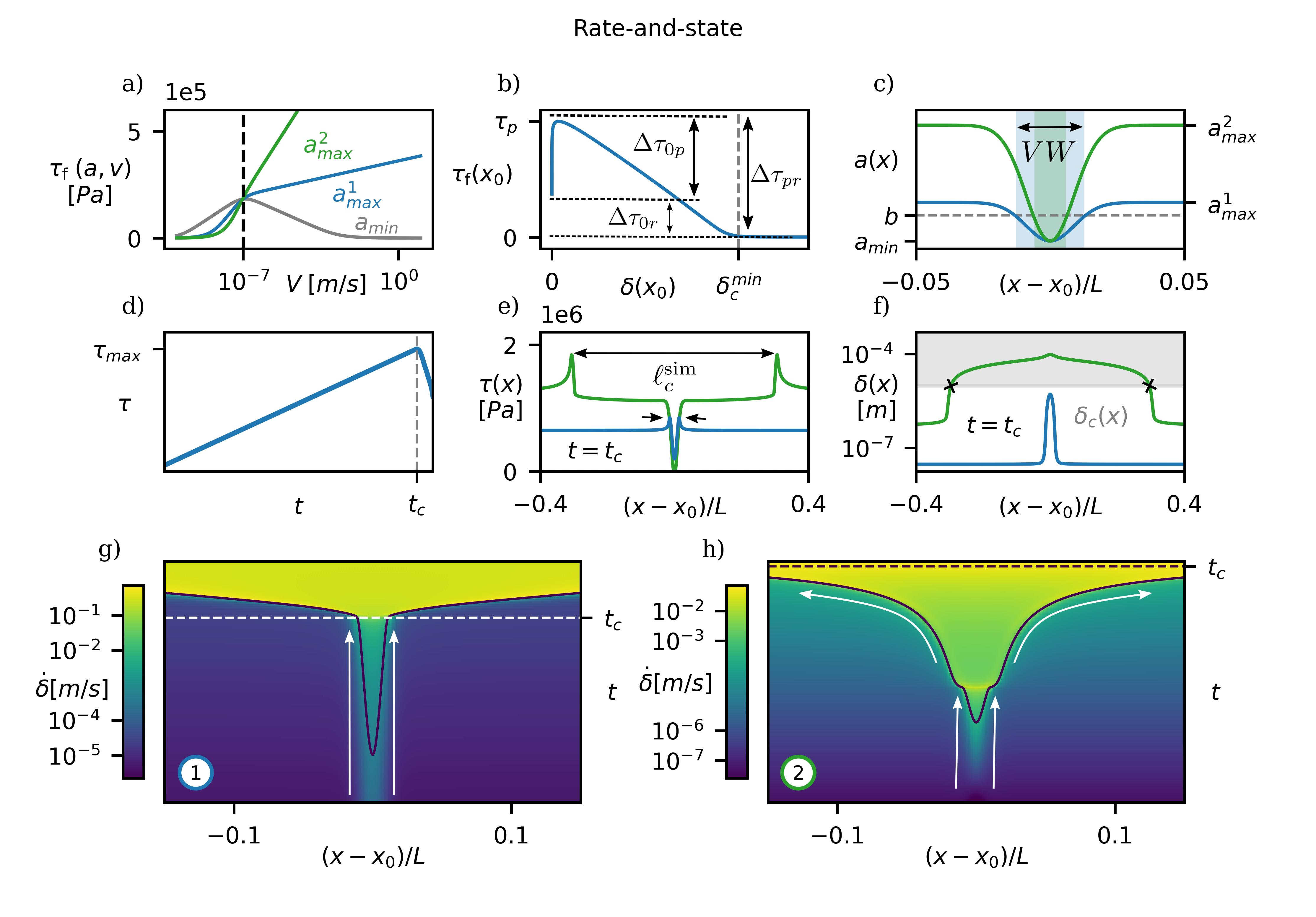}}
    \caption{Simulations of the onset of frictional sliding for a single nucleation patch governed by rate-and-state friction. a) Rate-and-state constitutive law for different values of $a$ in steady state ($\dot{\theta}=0$). b) Local shear stress evolution as a function of slip at $x=x_0$. $\Delta \tau_{0p}$ : Under-stress (difference between yielding and applied stress). $\Delta \tau_{0r}$ : applied-to-residual stress-drop. $\Delta \tau_{pr}$ : peak-to-residual stress-drop.  c) Initial configuration of rate-and-state constitutive parameters $a$ and $b$. The colored area marks the velocity-weakening (VW) patch, where $a>b$. d) Integrated macroscopic frictional stress over time. $t_c$ is defined as the $\tau(t=t_c) = \tau_\textrm{max}$. e) Shear stress distribution at the critical time. The measured nucleation length $\ell_c^{\textrm{sim}}$ is indicated by double black arrows. f) Slip distribution at the critical time. Shaded area represents $\delta(x) > \delta_c(x)$. $\delta_c(x)$ is approximated as $\delta$ such that $\tau_{\mathrm{f}}(\delta) \approx \tau_r$.  g,h) Slip-rate evolution in space and time. Dashed lines mark the critical time. Arrows indicate the localized phase and the expansion of the patch (h). }
    \label{fig:RS}
\end{figure}


 \begin{figure}
    \centering
    \includegraphics[scale=1]{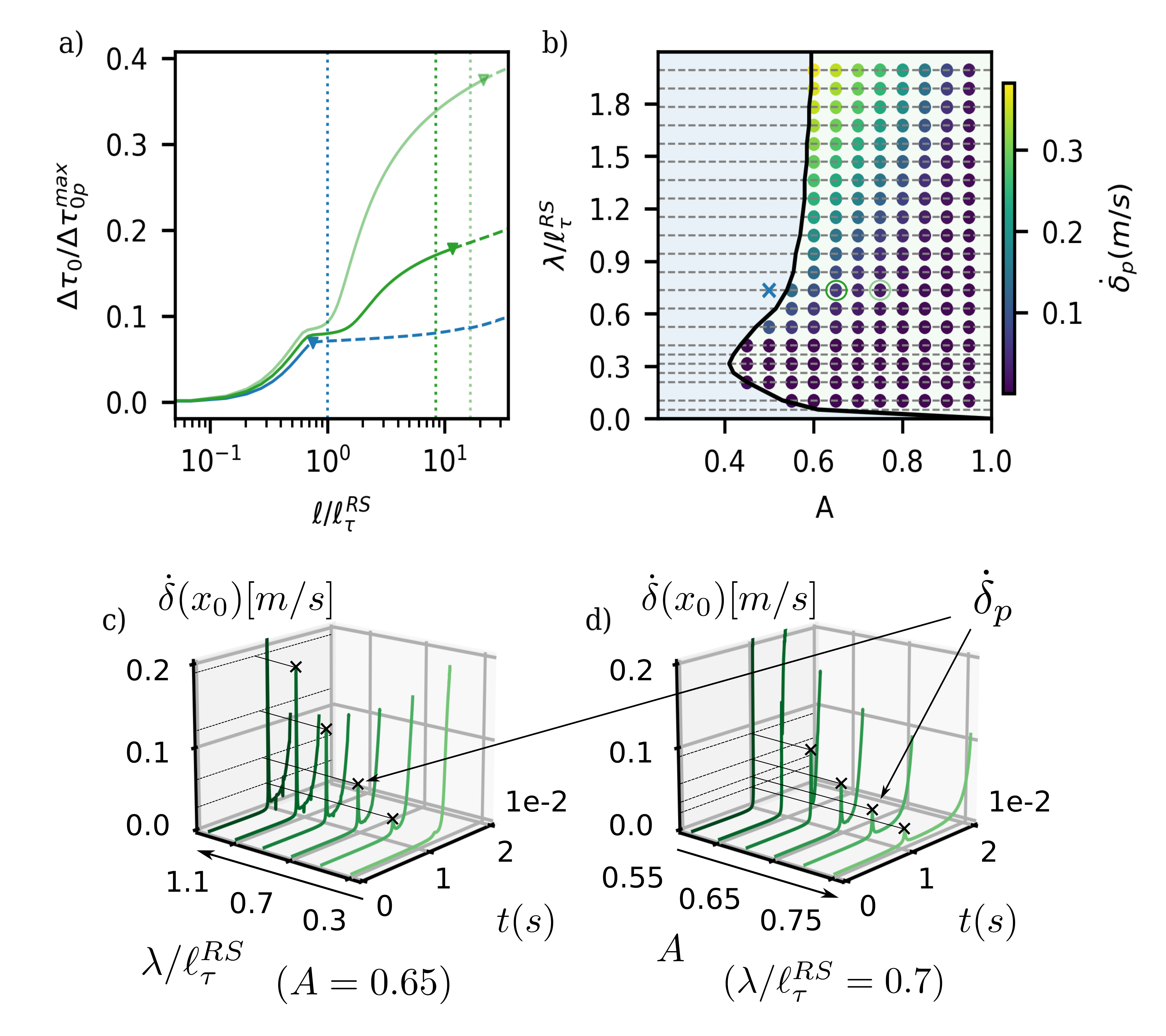}
    \caption{Nucleation regimes for a single nucleation patch governed by rate-and-state friction. \textbf{a)} Normalized load increment applied to the interface as a function of the normalized slip-patch length in log-scale. Colored triangles mark the critical point of instability. Vertical dotted lines indicate the analytical nucleation lengths as computed with Eq.~\ref{eq:RS-length} (in blue) and Eq.~\ref{eq:RS-crack-length} (in green), where $a = a_{min} + (b-a_{min})\times A$. \textbf{b)} Phase diagram showing the boundaries between the large-scale and small-scale yielding nucleation regimes for rate-and-state friction (black line). Dashed gray lines mark the values of $\lambda/\ell_{\tau}^{RS}$ for which the boundary values of $A$ were computed to draw the black line. Colored dots represent the maximal slip velocity of self-arrested dynamic events in the small-scale yielding regime $\dot{\delta}_p$. Blue '$\times$' and green circles refer to the simulations in '\textbf{a)}'. \textbf{c)} Slip-rate as a function of time at the weakest point of the interface, where slip nucleates, for $A=0.65$ and six different values of $\lambda / \ell_{\tau}^{RS}$. Velocity peaks $\dot{\delta}_p$ are marked with '$\times$'. \textbf{d)} Same as in '\textbf{c)}' for $\lambda/\ell_{\tau}^{RS} = 0.7$ and six different values of $A$.}
    \label{fig:RS-regimes}
\end{figure}

\subsection{Rate-and-state friction}
\label{sec:RSresults}

We further consider the case of an interface governed by rate-and-state friction. Here, the frictional strength $\tau_{\mathrm{f}}$ depends on the slip rate and behaves differently for different values of the rate-and-state constitutive parameter $a$. Specifically, the interface is velocity strengthening at all velocities in steady state ($\dot{\theta} = 0$) for $a>b$ (see Fig.~\ref{fig:RS}a), and velocity-weakening at $V > V_0$ for $a < b$. The under-stress or difference between the yielding and the applied stress $\Delta\tau_{0p}(x) = \tau_p(x) - \tau_0$, indicated on Fig.~\ref{fig:RS}b, is related to $a$ through $\Delta\tau_{0p}(x) = a \log(V/V_0)$~\cite{chenMicrophysicallyDerivedExpressions2017}. 
For this reason, we can create a weak nucleation patch by imposing a profile in $a(x) = q(x)$, similar to the $\tau_{p}(x)$ profile in the slip-weakening case. Specifically, this leads to a velocity-weakening patch where $b - a(x) > 0$ within a velocity-strengthening interface. We fix the minimum $a_{min} = 0.6 \times b$ and maintain $b$ constant. Hence, the velocity weakening patch shrinks when the amplitude of the profile increases (see Fig.~\ref{fig:RS}c), which will have an effect on stability during nucleation. We then apply a slowly increasing load and observe the behavior of the frictional interface. \par 

The overall behavior is similar to the slip-weakening case. The average interface traction increases until it suddenly fails at time $t=t_c$ (see Fig.~\ref{fig:RS}d). As previously observed, the failure occurs through a slowly growing yielding patch, which develops around the weak region (see Fig.~\ref{fig:RS}g), and eventually reaches a critical length and propagates dynamically across the interface.

We consider two interfaces with strength profiles of different amplitude (see Fig.~\ref{fig:RS}c). The two cases present important qualitative differences at the time of instability ($t=t_c$). The blue case with lower amplitude fails as slip is still confined in the velocity-weakening patch, which leads to a stress profile at $t=t_c$ (see Fig.~\ref{fig:RS}e) similar to the large-scale yielding regime in the slip-weakening case, and slip remains $\delta (x) < \delta_c (x)$ at any point (see Fig.~\ref{fig:RS}f; note that $\delta_c(x)$ varies weakly in space, so only its average is reported in the figure).  Therefore, the nucleation length can be approximated by Eq.\ref{eq:RS-length} and therefore, $\ell_c^{\textrm{sim}} \approx \ell_{\tau}^{RS}$, as seen on Fig.~\ref{fig:RS-regimes}a, where $C=0.77$ in Eq.~\ref{eq:RS-length} is taken from the experiments of McLaskey 2019~\cite{mclaskeyEarthquakeInitiationLaboratory2019}, while $a = a_{min}$ .

The green case with larger amplitude becomes unstable when slip already extends far into the velocity-strengthening zone. The stress profile at $t=t_c$, hence, presents an elongated area of relatively constant stress inside the slip patch (see Fig.~\ref{fig:RS}e) and $\delta (x) > \delta_c (x)$ in most parts of it (see Fig.~\ref{fig:RS}f). This nucleation is similar to the small-scale regime in the slip-weakening case, and it can be described by $\ell_c \approx \ell_{G}^{RS}$ (Eq.~\ref{eq:RS-crack-length}), where $a = a_{min} + (b-a_{min})\times A$ to account for the effect of the amplitude $A$. We note that $\ell_c^{sim} \approx \ell_{G}^{RS} \gg \ell_{\tau}^{RS}$, as expected~\cite{rubinEarthquakeNucleationAging2005}.

As already mentioned, the key difference between the two cases is that the pre-instability acceleration of slip is localized in the velocity-weakening patch for the low-amplitude case, and extends far beyond that for the high-amplitude case. In fact, the low-amplitude case becomes very abruptly unstable, whereas the high-amplitude case presents first a localized stage, which then evolves into a slowly accelerating crack that ultimately triggers the instability of the entire interface (indicated by curved-arrows in Fig.~\ref{fig:RS}h). These two phases of slip localization as opposed to crack-like expansion have been reported previously in similar numerical studies on nucleation~\cite{rubinEarthquakeNucleationAging2005,dublanchetDynamicsEarthquakePrecursors2018} as well as in laboratory experiments~\cite{ohnakaScalingShearRupture1999,mclaskeyEarthquakeInitiationLaboratory2019}, corroborating the existence of two nucleation phases. \par 

Finally, we explore again the parametric space for nucleation along a rate-and-state interface with a single weak patch. The overall properties are similar to those of the slip-weakening case. However, we only observe two qualitatively different nucleation regimes. There is the equivalent of the large-scale yielding regime, where the interface becomes unstable while still localizing (see blue case in Fig.~\ref{fig:RS-regimes}a); and, in addition, there is the regime with two phases (see two green cases in Fig.~\ref{fig:RS-regimes}a), where the patch transitions from a localized yielding area to an expanding crack until it fails. Unlike in the slip-weakening case, there is no qualitative difference between static and dynamic transition -- the transition is smooth but with varying degree of intensity. Considering the slip rate at $x=x_0$ (see Figs.~\ref{fig:RS-regimes}c\&d), we note a temporary acceleration. The maximal slip rate during this transition decreases with decreasing $\lambda$ and increasing $A$. \par

Exploring the full parametric space, we observe that the different nucleation regimes occupy similar parts to those in the slip-weakening case (compare Fig.~\ref{fig:RS-regimes}b with Fig.~\ref{fig:SW-regimes}b). While we cannot quantitatively distinguish between static and dynamic transition for the small-scale yielding regimes in the rate-and-state case, we do note that higher transitional velocity peaks $\dot{\delta}_p$ (see yellow region in Fig.~\ref{fig:RS-regimes}b), which point towards a more dynamic transition, occur where there was a dynamic transition in the slip-weakening case. Therefore, we can conclude that the slip-weakening and rate-and-state cases are fundamentally similar with differences in the abruptness of the static-vs-dynamic transition.

\begin{figure}
    \centering
    \centerline{\includegraphics[scale=1]{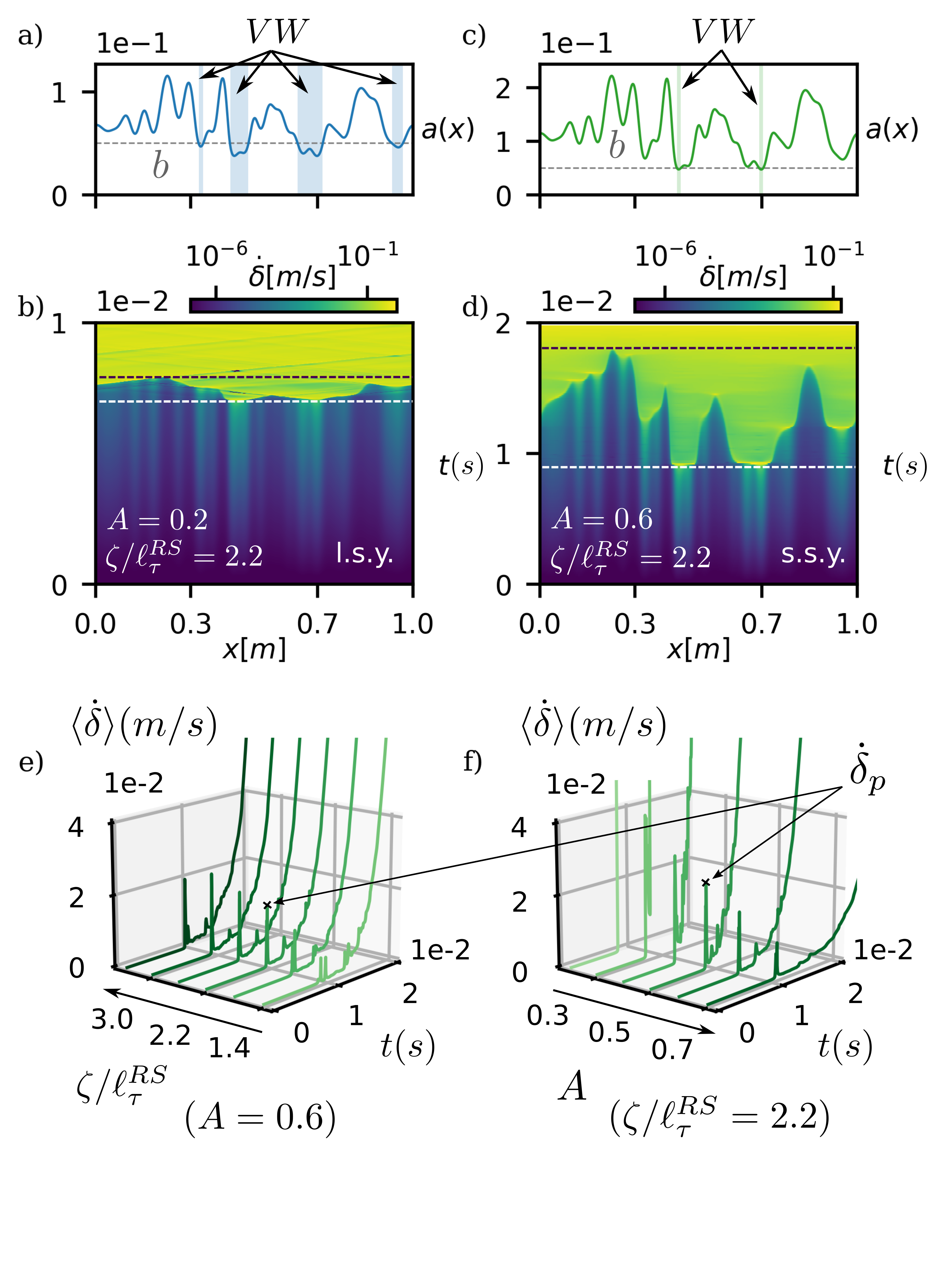}}
    \caption{ Nucleation on a random interface. \textbf{a,c)} Spatial distribution of frictional parameters 'a' (color) and 'b' (gray dashed). Velocity-weakening areas are shaded in color. \textbf{b,d)} Evolution of slip-rate in space-time for rate-and-state simulations of a stochastic interface with $\zeta/\ell_{\tau}^{RS} = 2.2$ and two different values of $A$. Dashed lines mark the failure of the first (white) and the last (dark) point. \textbf{e)} Averaged slip-rate over the interface as a function of time for an average amplitude of $A=0.6$ and six different values of correlation length $\zeta/\ell_{\tau}^{RS}$. \textbf{f)} Same as in 'e)' for $\zeta/\ell_{\tau}^{RS} = 2.2$ and six different values of $A$.}
    \label{fig:stochastic}
\end{figure}

\section{Nucleation on stochastic interfaces}
\label{sec:stochastic}

In the previous section, we studied the nucleation of frictional slip along interfaces with a simple weak patch. While this allowed us to observe and study different nucleation regimes, it is a simplistic system. Natural systems are more commonly characterized by multiple and random heterogeneities created by surface roughness and material microstructures, which is expected and observed~\cite{mclaskeyEarthquakeInitiationLaboratory2019,scharNucleationFrictionalSliding2021} to lead to richer dynamics during nucleation. Here, we aim to translate our findings from single weak-patches to more realistic stochastic frictional interfaces.

We use a rate-and-state interface where we impose a distribution of $a(x)$ that we generate using the algorithm from Albertini et. al. 2021~\cite{albertiniStochasticPropertiesStatic2021}, which allows to define the under-stress $\Delta \tau_{0p}(x)$ along the interface with a given amplitude $A = a_{max} - a_{min}$ and a correlation length $\zeta$ (see Fig.~\ref{fig:stochastic}a\&c).  We choose the value of $b$ to be above the minimum of $a(x)$ ($a_{min} = 0.6 \times b$) to ensure a few velocity-weakening patches. A higher b-value would lead to many velocity weakening patches, making the interface too unstable, whereas a value of b below $a_{min}$ would hinder nucleation. We then apply a uniform quasi-statically increasing load to the interface, which leads to the nucleation of multiple slip patches along the weakest points of the interface (compare Figs.~\ref{fig:stochastic}a\&e to \ref{fig:stochastic}b\&d, respectively). These slip patches continuously grow -- in some cases occasionally accelerating and decelerating -- until the entire interface accelerates and produces global sliding. 

Can we observe the same nucleation regimes as in the single weak-patch configuration? 
The first representative case shown in Fig.~\ref{fig:stochastic}a\&b, clearly resembles the large-scale yielding regime (l.s.y.) of the single patch configuration (see Sec.~\ref{sec:RSresults}) as the entire interface fails almost simultaneously. This is confirmed by the slip rate that grows instantaneously and rapidly (see case $A = 0.2$ in Fig.~\ref{fig:stochastic}f). The second representative case (see Fig.~\ref{fig:stochastic}c\&d), however, presents a failure mechanism that is considerably slower (see dashed lines in Fig.~\ref{fig:stochastic}b\&d), and that is more similar to the small-scale yielding regime (s.s.y.). In this case, the interface first breaks at the weaker points, triggering localized slip events of higher slip rate (see yellow parts in Fig.~\ref{fig:stochastic}d), and then fails through the coalescence of multiple slowly growing cracks. These local slip events are created by the transition from yielding to fracture around the weakest points, which, for rate-and-state stochastic interfaces, is usually dynamic and causes peaks in slip rate $\dot{\delta}_p$ (see Fig.~\ref{fig:stochastic}e and f for $A>0.3$). We observe that $\dot{\delta}_p$ increases with longer $\zeta$ and decrease with larger $A$ (see Figs.~\ref{fig:stochastic}e\&f). 

Lastly, comparing the size of the velocity-weakening patches in both cases (shaded regions in Figs.~\ref{fig:stochastic}a\&c), it seems that it could be an important factor contributing to the abruptness of instability in the large-scale yielding regime.

All in all, these results show that both nucleation regimes, \textit{i.e.}, large-scale yielding and small-scale yielding, exist at stochastic frictional interfaces, and that smaller amplitudes in the heterogeneity favor the large-scale yielding failure regime. 

\section{Discussion} 
\label{sec:five}

In this section, we discuss the implications of our findings for various applications involving frictional interfaces. For instance, foreshock activity is often considered to be the cause and/or result of the earthquake nucleation process along natural faults~\cite{rubinsteinDynamicsPrecursorsFrictional2007,popovAcceleratedCreepPrecursor2010,mclaskeyForeshocksNucleationStickslip2013,costagliolaCorrelationSlipPrecursors2022a}. Our results suggest that some of these foreshocks and other precursory events could be the frictional instabilities resulting from unstable transitions between nucleation phases (from yielding to fracture). In fact, a recent study~\cite{weiSlowSelfarrestingNature2021} suggests that low-frequency earthquakes (LFEs) are slow self-arresting ruptures generated by the failure of small brittle asperities within stronger and more ductile environments of the kind found at the deeper end of the subduction zone, where these events are registered~\cite{passarelliSourceScalingSeismic2021}. According to our model, this can be approximated by a system with low correlation length (small asperities) and large amplitude, where unstable transitions from yielding to fracture would indeed produce relatively slow but dynamic self-arresting events, matching the signature of LFEs. \par 

Other works have studied LFEs in the context of slow-slip events (SSEs)~\cite{frankSlowSlipHidden2016,frankDailyMeasurementSlow2019}, which are long-lived events that can span several kilometers along subduction zones. According to these works, LFEs very often accumulate on the down-dip edge of slow slipping patches~\cite{liuRecurrentSlowSlip2015,chestlerModelLowFrequencyEarthquake2017}, suggesting that conditions at the boundary of these patches are particularly favorable for LFEs. 
In our model, large-scale asperities lead to large areas of slow slip (being already in the fracture phase) with localized yielding along the boundary (see small-scale yielding regime in rate-and-state case). Within this localized area, sub-asperities undergo a transition from yielding to fracture, which may generate seismic activity in the form of tremor or  swarms~\cite{passarelliSourceScalingSeismic2021}. Hence, our model provides an explanation for the localization of seismicity at the boundary of slow slipping areas.  \par 

Nucleation of frictional sliding is often studied in large-scale laboratory experiments. Recent work by Gvirtzman and Fineberg, 2021~\cite{gvirtzmanNucleationFrontsIgnite2021} showed that nucleation in their experiments is governed by a Griffith's type of critical length (\textit{i.e.}, small-scale yielding regime), which is in contrast to the often made assumption of a large-scale yielding regime nucleation in earthquake mechanics studies~\cite{tapeEarthquakeNucleationFault2018}. Our work provides a possible explanation for this discrepancy. In the experiments~\cite{gvirtzmanNucleationFrontsIgnite2021}, the stress state at the point of nucleation is highly concentrated\footnote{It is the result of an arrested crack and hence corresponds to the near-tip stress concentration/singularity.}, \textit{i.e.}, high amplitude and small correlation length, which, according to our findings (see Fig.~\ref{fig:SW-regimes}b), corresponds to the small-scale yielding regime likely with static transition. Natural faults, however, might be considerably smoother -- this remains to be proven -- with lower amplitude and larger correlation length, which would suggest that the large-scale yielding regime is more relevant for earthquakes.

Lastly, our results are also relevant for frictional interfaces in engineering applications (\textit{i.e.}, at lower scales). For instance, in the field of topologically interlocked materials and structures~\cite{mirkhalaf2018}. These structures are formed by the geometrical interlocking of building elements without any adhesive components but simply through friction. Therefore, their strength relies heavily on the bearing capacity of frictional interfaces.  Recent work~\cite{koureasFailureBeamlikeTopologically2022} has shown that sliding has a crucial impact onto the performance of such structures and hence, understanding and predicting the nucleation of slip along these interfaces is key to better design approaches. Our work provides the necessary fundamental understanding for such future developments. For example, tuning the frictional heterogeneity of these interfaces would allow to build structures that fail gradually instead of abruptly, which would make them safer.

\section{Conclusion}
\label{sec:six}
We have demonstrated that the collective statistical features of friction along a given interface play a decisive role in determining the nucleation mechanism of frictional instabilities. In particular, we have shown that large frictional heterogeneities unlock the onset of residual friction during nucleation, which can have a significant impact on the nucleation dynamics of slip. We were able to identify three different nucleation regimes, whose occurrence depends on the correlation length and the amplitude of frictional heterogeneities. These nucleation regimes rely on the stability of two different phases and the interplay between them; A yielding phase, driven by the weakening of asperities inside the slip-patch, and a fracture phase, governed by residual friction. Moreover, we found a mechanism responsible for local dynamic events during nucleation as a result of unstable transitions between these two phases. Our work shows the importance of local failure mechanisms in the nucleation of collective events and stresses the role of heterogeneity through amplitude and correlation length in the failure dynamics of frictional interfaces, which can help us to better understand the nature of seismic activity in natural faults but also to design better structures. 

\section*{Acknowledgements}

The simulation data generated in this study have been deposited in the ETH Research Collection database available at https://doi.org/10.3929/ ethz-x-xxxxxxxx.

\appendix

\section{Piece-wise Constant Slip Method}
\label{sec:slipmethod}

This method, which was adapted from Garagash and Germanovich, 2012 ~\cite{garagashNucleationArrestDynamic2012} by removing the terms coming from pore-pressure, is used to solve the static equilibrium equation (Eq.~\ref{eq:quasi-static}) over a discrete domain, allowing us to compute the equilibrium values of the external load ($\tau_0$) for any given size of the slip-patch ($2d = |\ell_+ - \ell_-|$). To do this, we discretize the slip-patch domain into $2N + 1$ uniformly spaced elements ($\Delta X = 1/N$) with midpoints $X_j = j\Delta X (j = -N, ..., 0, ..., N)$, such that $X = (x-x_0)/d \in [-1,1]$ is the normalized coordinate along the slip-patch. This discretization turns Eq.~\ref{eq:quasi-static} into
\begin{equation}
    \tau_\mathrm{f}(i\Delta X) = \tau_0 - \frac{\mu^{\prime}}{d}\sum_j K_{ij}\delta(j\Delta X)
\end{equation}
where $K_{ij} = -1/(2\pi \Delta X [(i-j)^2 -1/4])$ and $\delta(\pm N\Delta X) = 0$, which leaves us with a total of $2N+1$ equations for $2N+1$ slip unknowns ($\delta(j\Delta X)$) plus the external load $\tau_0$. Therefore, we need an extra equation, which we draw from the finiteness conditions, stating 
\begin{equation}
    \int\displaylimits_{\ell_-}^{\ell_+} \frac{\tau_\mathrm{f}(x)-\tau_0}{\sqrt{d^2 -x^2}}dx = 0, \; \; \int\displaylimits_{\ell_-}^{\ell_+} \frac{\tau_\mathrm{f}(x)-\tau_0}{\sqrt{d^2 -x^2}}xdx = 0
\end{equation}

The second equation is automatically satisfied by symmetry of the slip-patch, and the discretization of the first one yields
\begin{equation}
    \tau_0 = \sum_j k_j \tau_\mathrm{f}(j\Delta X)
\end{equation}
where 
\begin{equation}
    k_j = \frac{1}{\pi} \int\displaylimits_{\text{max}\{-1,(j-0.5)\Delta X\}}^{\text{min}\{+1,(j+0.5)\Delta X\}} \frac{dX}{\sqrt{1-X^2}}.
\end{equation}
With this last equation, we have $2N+2$ equations for $2N+2$ unknowns so we can solve the linear system for any given half-length $d$. For the static solutions displayed on Fig.~\ref{fig:SW-regimes}a of the manuscript, we used $N=30$. However, for the solutions used to calculate the critical lengths in Fig~ \ref{fig:SW-regimes}d, $N=80$ was used. 

\newpage
\section{Simulation parameters}
\label{sec:simparams}

The parameters applied in our simulations are summarized in Table~\ref{tab:tab1}.

\begin{table}[h]
\begin{center}
\begin{tabular}{||c c c||} 
 \hline
 \multicolumn{3}{||c||}{Elastic} \\
 \hline
 Description & Parameter & Value  \\ 
 \hline\hline

 Poisson's ratio & $\nu$ & 0.33 \\
 Density & $\rho$ & 1170 $kg/m^3$ \\
 Young's modulus & $E$ & 2.65 GPa \\
  \hline
  \multicolumn{3}{||c||}{Rate-and-state} \\
  \hline
 Description & Parameter & Value  \\ 
 \hline\hline
  Contact pressure & $\sigma$ & $-1$ MPa \\
  Characteristic slip & $D_c$ & $5\times 10^{-7}$ m \\
  Reference velocity & $V^*$ & $10^{-7}$ m/s \\ 
  Constitutive parameter & $b$ & 0.05 \\
  Constitutive parameter & $a_{min}$ & $0.03$ \\
  Reference friction coefficient & $\mathrm{f}^*$ & 0.2 \\
  Shape factor & $C$ & 0.77 \\
 \hline
  \multicolumn{3}{||c||}{Slip-weakening} \\
  \hline
 Description & Parameter & Value  \\ 
 \hline\hline
 Peak-strength & $\tau_p^{max}$ & $1$ MPa \\ 
 Weakening rate & $W$ & $10^{6}$ MPa/m\\ 
 Residual strength & $\tau_{r}$ & 0 Pa\\ 
 \hline
\end{tabular}
 \caption{Fixed parameters for rate-and-state and slip-weakening simulations.}
\label{tab:tab1}
\end{center}
\end{table} 

 \bibliographystyle{elsarticle-num} 
 
 \bibliography{references,references2}





\end{document}